\documentclass[12pt]{iopart}
\usepackage{graphicx}

\begin{document}
\title[Mohanty et. al]{Observation of Topological Hall Effect and Skyrmions in Pt/Co/Ir/Co/Pt System}

\author{Shaktiranjan Mohanty$^{1}$, Brindaban Ojha$^{1}$, Minaxi Sharma$^{1,2}$ and Subhankar Bedanta$^{1}$}

\address{$^{1}$Laboratory for Nanomagnetism and Magnetic Materials (LNMM), School of Physical Sciences, National Institute of Science Education and Research (NISER), An OCC of Homi Bhabha National Institute (HBNI), Jatni, Odisha 752050, India.}
\address{$^{2}$School of Basic Sciences, Department of Physics, Bahra University Waknaghat, Solan, 173234, H. P., India.}
\ead{sbedanta@niser.ac.in}
\vspace{10pt}
\begin{abstract}
The interlayer exchange coupling (IEC) between two ferromagnetic (FM) layers separated by a non-magnetic (NM) spacer layer gives rise to different types of coupling with the variation of spacer layer thickness. When the NM is metallic, the IEC is attributed to the well-known Ruderman–Kittel–Kasuya–Yosida (RKKY) interaction which shows an oscillatory decaying nature with increasing thickness. Due to this, it is possible to tune the coupling between the two FM to be either ferromagnetic or antiferromagnetic. In this work we have studied a Pt/Co/Ir/Co/Pt system where the Co thickness has been taken in the strong perpendicular magnetic anisotropy regime which is much less than the spin reorientation transition thickness. By tuning the Ir thickness to 2.0 nm, a canted state of magnetization reversal in the system is observed which gives rise to a possibility of nucleating topologically non-trivial spin textures like skyrmions. Further, with the combination of transport and magnetic force microscopy (MFM) measurements, we have confirmed the presence of skyrmions in our system. These findings may be useful for potential applications in emerging spintronic and data storage technologies using skyrmions.
\end{abstract}
\section{Introduction}

Skyrmions are  highly promising topological structures that have garnered significant attention due to potential applications in spintronic devices such as magnetic data storage technologies and logic devices\cite{luo2018reconfigurable,dash2023device,zhang2015magnetic}. They were first introduced in the context of theoretical nuclear physics by the British physicist Tony Skyrme in the 1960s\cite{skyrme1961non}, and later they are found in the fields of condensed matter physics in particular in the field of magnetism\cite{bogdanov1989thermodynamically,bogdanov1994thermodynamically,bogdanov2020physical,muhlbauer2009skyrmion,leonov2016properties}. Skyrmions are quasiparticles that manifest as localized, swirling patterns of magnetic moments or spins in a material. These swirling structures are stabilized by competing interactions/energies within the material such as exchange interaction, anisotropy\cite{moreau2016additive,boulle2016room}, Dzyaloshinskii-Moriya interaction (DMI)\cite{rohart2013skyrmion}, frustrated exchange interaction\cite{von2017enhanced}, dipolar interaction\cite{ezawa2010giant} etc. The unique topological characteristics of skyrmions make them distinct from other magnetic configurations and offer several advantages such as small size, low energy consumption, stability and potential for race track memory application\cite{fert2013skyrmions,fert2017magnetic,nagaosa2013topological,tomasello2014strategy,zhang2015skyrmion,sampaio2013nucleation}. Apart from these, skyrmions are also very interesting from fundamental physics point of view. However, creating a good platform for the deterministic nucleation as well as detecting the presence of these quasi particles in the system through simplest techniques available is still a crucial part of this field. 
Previously, skyrmions have been observed in metallic thin films consisting of single FM layer with thickness near to spin reorientation transition (SRT)\cite{ojha2023driving}, in multilayer system with significant repetition of the FM and NM layers\cite{chen2022controllable,dohi2019formation,chen2020realization,legrand2020room} etc. Raju et. al., have observed high density skyrmions by taking multilayers of Ir/Fe/Co/Pt system by enhancing the DMI at both the interface of Ir/Fe and Pt/Co\cite{raju2019evolution}. Similarly, Chen et. al., have observed the presence of skyrmions in Pd/Co multilayers with Ru spacer\cite{chen2020realization}. In the later work, the competition between DMI, dipolar interaction as well as RKKY interaction leads to the formation of these chiral spin textures.  However, there are no such reports where skyrmions have been observed just by varying the NM spacer layer thickness between two FM layers having thickness in perfectly PMA regime i.e. far away from the SRT regime.
In this paper, we focus on a comprehensive exploration of the THE and skyrmions in a Pt/Co/Ir/Co/Pt system, with particular emphasis on how varying the Ruderman-Kittel-Kasuya-Yosida (RKKY) interaction through the Ir spacer layer affects these phenomena. The RKKY interaction\cite{bruno1991oscillatory,bruno1992ruderman}, a long-range indirect exchange interaction mediated by conduction electrons, plays a crucial role in determining the magnetic properties of multilayered systems.  By manipulating the strength and nature of the RKKY interaction via the thickness of the Ir spacer layer, we gain insight into how it influences the emergence and stability of skyrmionic structures, as well as the corresponding Topological Hall Effect signals.
In our previous work, we have explored the magnetization reversal and domain structures in the same system by varying the Ir thickness as well as the number of Co layers\cite{mohanty2022magnetization}. There we have varied the Ir spacer layer thickness as 1.0, 1.5 and 2.0 nm between two Co layers of 0.8 nm thickness. We found that for Ir 1.0 nm the interlayer exchange coupling (IEC) is ferromagnetic (FM) and for Ir 1.5 nm, the IEC is antiferromagnetic (AFM) i.e., synthetic antiferromagnets (SAF)\cite{duine2018synthetic}. In both of these multilayers with Ir thickness of 1.0nm and 1.5nm, we have observed bubble domains during the magnetization reversal which can be attributed to the higher anisotropy in the system. However, for the multilayer with Ir thickness of 2.0 nm, the hysteresis loop showed a slanted type of reversal which indicates that the magnetizations of different Co layers are in a relatively canted state during the reversal. Kerr microscopy results show very small ripple like domains images with the higher objective which is already shown in the previous work\cite{mohanty2022magnetization}. In order to gain deeper insight to the magnetic structure in this work we have considered magnetic force microscopy for imaging. Further topological Hall effect experiments are performed to confirm the observed magnetic microstructure to be chiral in nature. 
\section{Experimental Details:}
Two multilayer samples have been prepared in a high-vacuum multi-deposition chamber manufactured by Mantis Deposition Ltd., UK. The samples are named as Ir-2-1 and Ir-1-1 having layer structure as Si/Ta(3)/[Pt(3.5)/Co(0.8)]2/Ir(2.0)/Co(0.8)/Pt(3.5) and Si/Ta(3)/Pt(3.5)/Co(0.8)/Ir(2.0)/Co(0.8)/Pt(3.5),  respectively The base pressure of the chamber was better than 1$\times$10$^{-7}$ mbar. The deposition pressure was $\sim$ 1.5$\times$10$^{-3}$ mbar for Ta and Pt layers. Furthermore, for Co and Ir, the deposition pressures were $\sim$ 5$\times$10$^{-3}$ and $\sim$ 2.1$\times$10$^{-3}$ mbar, respectively. During sample preparation, the substrate table was rotated at 15 rpm to minimize the growth-induced anisotropy and also to have uniform growth of the films. 
The rates of deposition were 0.1$\AA$/s, 0.13$\AA$/s, 0.3$\AA$/s and 0.1$\AA$/s for Ir, Ta, Pt and Co, respectively. 
We have performed the hysteresis measurements at room temperature using a SQUID-VSM (Superconducting Quantum Interference Device-Vibrating sample magnetometer) MPMS3 manufactured by Quantum Design, USA to measure the magnetization of the sample. Magneto transport measurement has been performed with a physical property measurement system (PPMS) by Quantum Design, USA, to extract the THE signal of the sample. A small rectangular sample was applied with a constant current of 0.5 mA along the longitudinal direction for measuring the resistivity of the sample and the voltage was measured along the transverse direction as shown in the figure \ref{fig:fig3} (b). Presence of skyrmions have been observed via magnetic force microscopy (MFM) measurement using Attodry 2100 AFM/MFM system manufactured by Attocube, Germany.
\section{Result and discussion:}
\begin{figure}
	\centering
	\includegraphics[width=1\textwidth]{"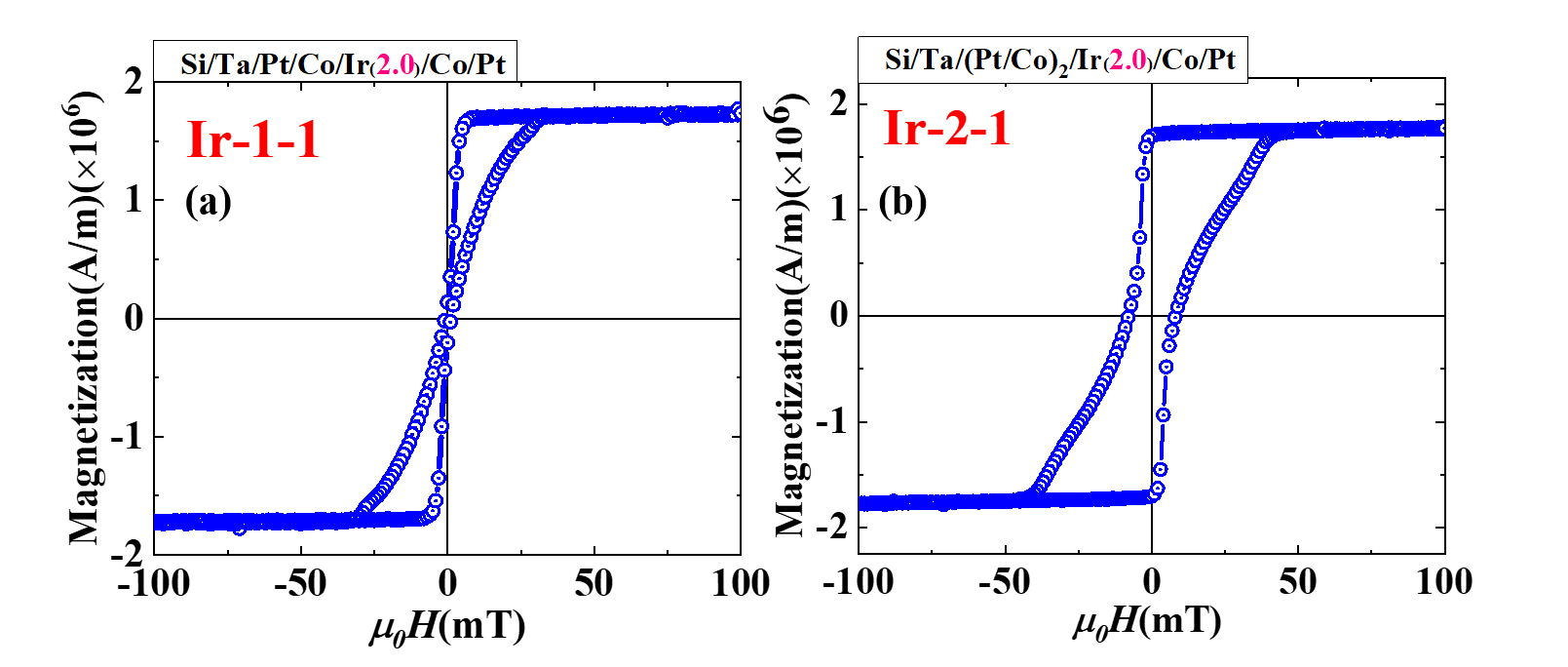"}
	\caption{Out-of-plane magnetic hysteresis loops of sample (a) Ir-1-1 and (b) Ir-2-1 measured by SQUID-VSM.}
	\label{fig:fig1}
\end{figure}

The schematic of the samples Ir-1-1 and Ir-2-1 has been shown in the figure S1 of the supplementary information. In our previous work[29], we have shown the transmission electron microscopy (TEM) on another sample of this series having Co thickness 0.8 nm and the Ir thickness 1.5 nm. The continuous growth of these ultrathin layers has been observed from the TEM images\cite{mohanty2022magnetization}. In order to understand the detailed structural property of the multilayer stack presented in this work, we have performed XRR for sample Ir-1-1. The fitting of the XRR data for Ir-1-1 sample is shown in Fig. S2 of the supplementary information. The thicknesses of the layers are found to be 3.31, 3.73, 0.83, 2.05, 0.79, 3.08 nm for Ta, Pt, Co, Ir, Co, Pt layers, respectively. Also, the roughnesses of the layers are found to be 0.29, 0.49, 0.53, 0.31, 0.59, 0.36, respectively. The error bars for each layers are mentioned in the supplementary information table-1.
Next we show the magnetic characterization of the samples. The magnetization reversal of the samples Ir-1-1 and Ir-2-1 have been shown in Figure \ref{fig:fig1}(a) and (b), respectively. It is observed that the magnetic hysteresis loops showed an out-of-plane easy axis with slanted magnetization reversal. The small remanence at zero magnetic field for Ir-1-1 is due to the balance of up and down domains with one layer of Co below and above the Ir spacer. However, for Ir-2-1, as there are two Co layers below the Ir and one Co layer above the Ir, a finite remanence is observed at zero field in this sample due to the imbalance of up- and down-domains\cite{salikhov2022control}. It should be noted that the other samples of this series having Ir thickness 1.0 nm showed a sharp reversal in the out-of-plane applied magnetic field which indicates the FM coupling between the two Co layers. Whereas by changing the Ir thickness to 1.5 nm, the coupling became AFM leading to SAF structure. Further, bubble domains have been observed in these samples having Ir thickness 1.0 nm as well as 1.5 nm\cite{mohanty2022magnetization}. However, the samples Ir-1-1 and Ir-2-1 show small ripple domains during the magnetization reversal as shown in the Fig. S4 of the supplementary information (with permission from Springer Nature). Due to the limitations (resolution) of our Kerr microscopy, we could not properly investigate these observed very small ripple domains. However, the loop shape as well as the domain structures indicated the possible presence of chiral spin textures like skyrmions in it. 

\begin{figure}
	\centering
	\includegraphics[width=1\textwidth]{"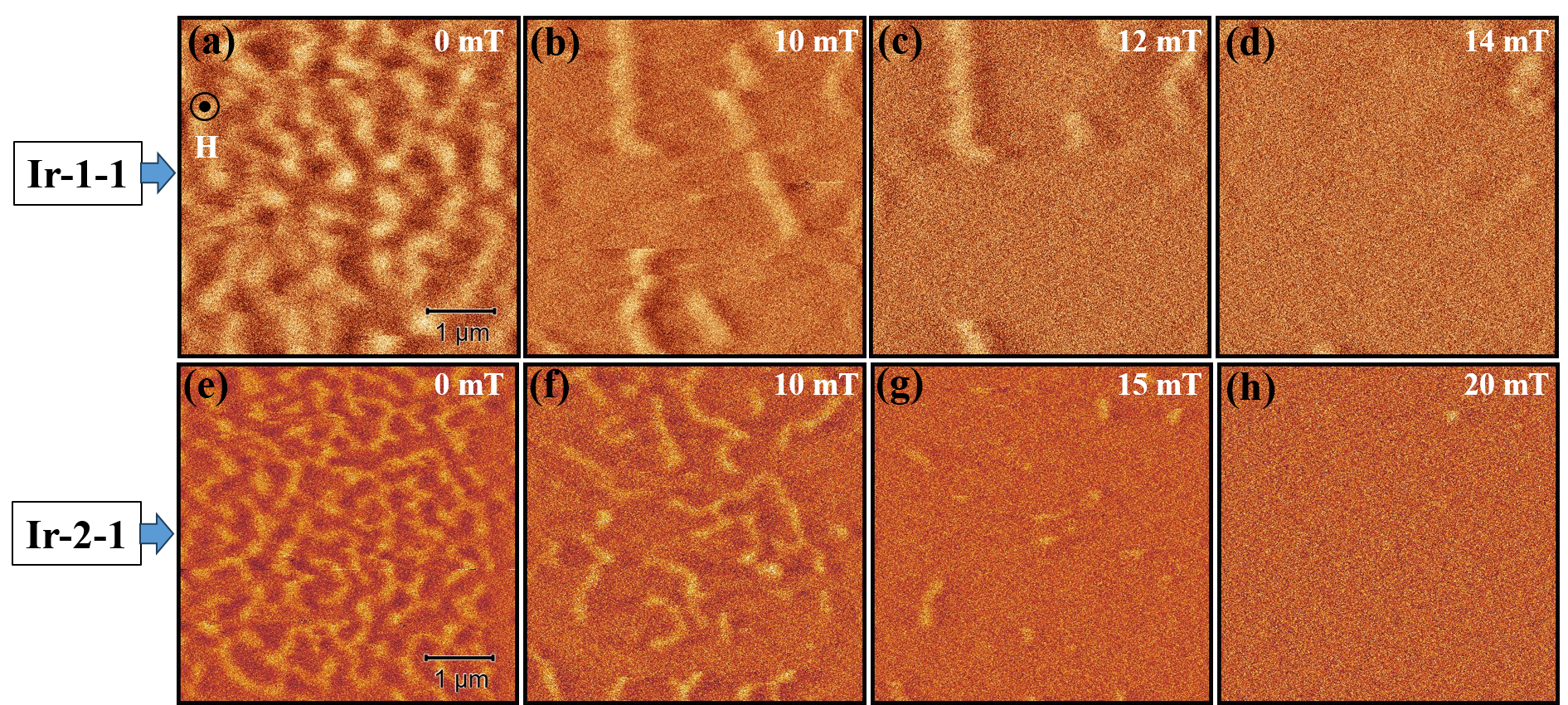"}
	\caption{(a)-(d) are the MFM images of the sample Ir-1-1 at different external applied field of 0 mT, 10 mT, 12 mT and 14 mT, respectively. (e)-(h) are the MFM images of the sample Ir-2-1 at applied field 0 mT, 10 mT, 15 mT and 20 mT, respectively The scale bar of 1 µm is shown (a) and (e).}
	\label{fig:fig2}
\end{figure}

In order to observe if there is existence of any skyrmioninc states, one needs to perform high resolution magnetic imaging. There are several methods to detect the presence and characteristics of skyrmions in magnetic materials. A few can be pronounced here such as, magnetic force microscopy (MFM)\cite{ojha2023driving}, magneto optic Kerr effect (MOKE) microscopy\cite{chen2022controllable,dohi2019formation},  scanning tunneling microscopy (STM)\cite{palotas2017spin}, X-ray photoemission electron microscopy (XPEEM)\cite{juge2022skyrmions}, Hall Effect Measurements, Lorentz transmission electron microscopy (LTEM)\cite{chen2020realization,he2018evolution} etc. In this work, we have performed the magnetic force microscopy (MFM) measurement on both the samples Ir-1-1 and Ir-2-1 to detect the presence of skyrmions. The MFM images in figure \ref{fig:fig2} (a) and (e) are taken at the demagnetized state (i.e., without applying any external magnetic field to the sample) for samples Ir-1-1 and Ir-2-1, respectively. In this condition we have observed labyrinth kind of domains or spin spirals which may be the outcome of the dominance of DMI, which favors the non-collinear spin textures, over the exchange favoring the collinear spin textures at this state\cite{ojha2023driving,raju2019evolution,raju2021colossal}. By minimizing the total energy which includes exchange, DMI, anisotropy etc., one can get the spin spiral states in a system at demagnetized state. The total energy of the system can be written as as shown in equation \ref{eq:Energy},
\begin{equation}
    E_{tot} = E_{exch} + E_{DMI} + E_{anis} + E_{Zeem} + E_{RKKY}
    \label{eq:Energy}
\end{equation}
Where, $E_{exch}$ is the exchange energy, $E_{DMI}$ is the DMI energy, $E_{anis}$ the anisotropy energy, $E_{Zeem}$ is the Zeeman energy and $E_{RKKY}$ is the RKKY interaction energy. This Zeeman energy tends to align the magnetic moments along the field direction. It leads to the nucleation of skyrmions. Hence, further applying external magnetic fields in both the samples, these labyrinth domains gradually break into smaller part and form chiral spin texture like skyrmions. By further applying a higher magnetic field, the skyrmions get annihilated and a uniform saturated magnetic state is formed. It is found that the size of the skyrmions in Ir-2-1 is smaller as compared to the size of skyrmions in Ir-1-1. The reason may be attributed to the comparatively smaller anisotropy energy of sample Ir-2-1 which favors the smaller size of skyrmions[24]. The values of $K_{eff}$  are found to be 5.55 $\times$ 10$^5$  and 7.44 $\times$ 10$^5$  J/m$^3$  for samples Ir-1-1 and Ir-2-1, respectively\cite{mohanty2022magnetization}.

Further to confirm the chirality of these observed spin textures, we have performed the magneto-transport measurements on a small rectangular piece of the sample Ir-1-1 in the Van der Pauw geometry as shown in the supplementary figure S5 (a). The THE, characterized by the generation of a transverse charge current perpendicular to an applied magnetic field, has been a subject of extensive investigation due to its deep connection to the underlying magnetic texture's topological properties and becomes even more important when coupled with the existence of skyrmions. The conduction electrons' interactions with the chiral spin textures like skyrmions result in an emergent magnetic field, which is what causes the THE\cite{he2018evolution,soumyanarayanan2017tunable}. In such magneto-transport experiments three effects have been considered which contribute to the total Hall resistivity of the sample. The ordinary Hall effect (OHE) that has linear contributions in accordance with the magnetic field. The anomalous Hall effect (AHE) which usually arises in a FM due to the Berry curvature in momentum space, scales linearly with the perpendicular component of magnetization. Then the THE related to the Berry phase in real space, occurs due to the presence of topologically nontrivial or chiral spin textures like skyrmions in the system. Electrons when move in such chiral spin textures, experiences an extra fictitious magnetic field, gets deflected perpendicular to the current direction giving rise to the THE\cite{raju2019evolution,chen2020magnetic,sivakumar2020topological,kimbell2022challenges}. Hence, the total Hall resistivity of the sample can be written as equation \ref{eq:Rho},
\begin{equation}
    \rho_{xy} = R_0 H +R_s M + \rho_{THE}
    \label{eq:Rho}
\end{equation}

\begin{figure}
	\centering
	\includegraphics[width=1\textwidth]{"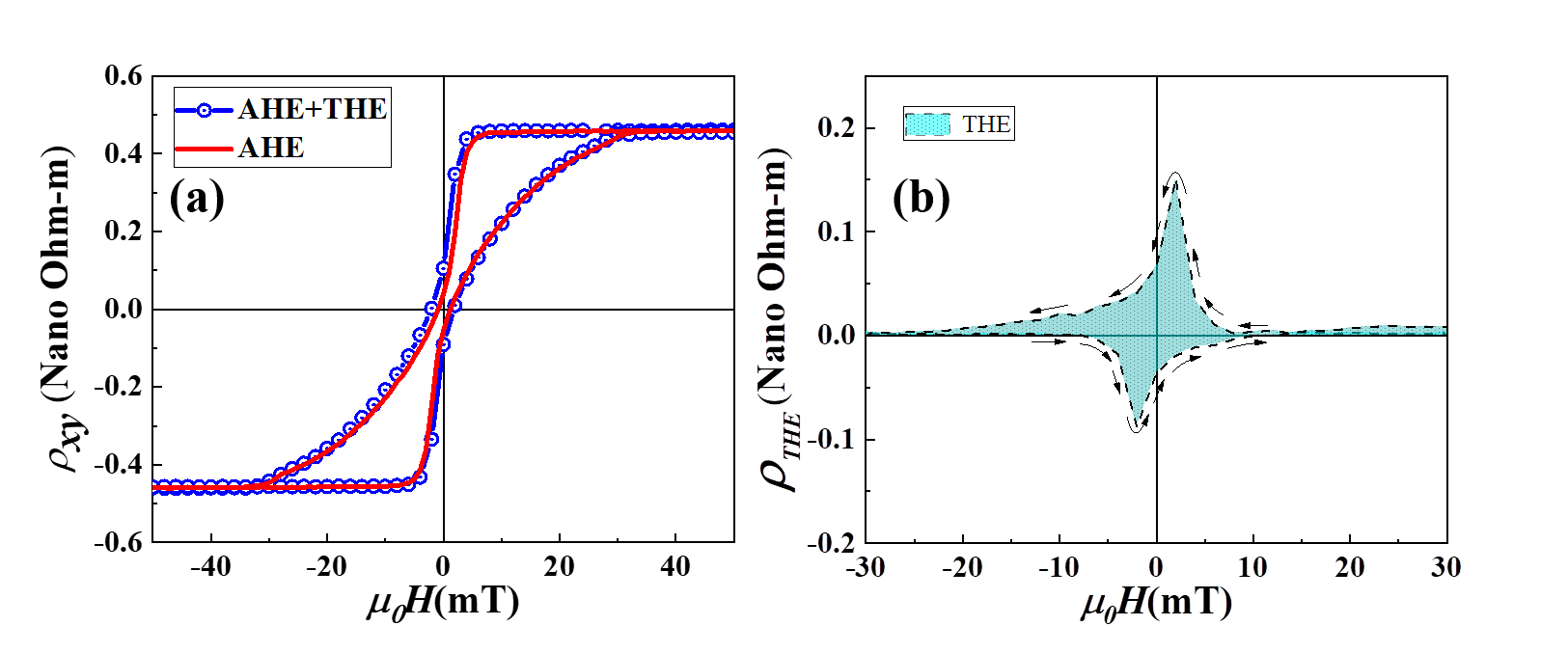"}
	\caption{Topological Hall effect (THE) measurement of the sample Ir-1-1. (a) Total resistivity of the sample after removing the ordinary Hall effect component (blue circles) and the AHE fitting to the total Hall resistivity (red line), (b) $\rho_{THE}$ extracted from the resistivity vs magnetic field measurement data after subtracting the OHE and AHE part.}
	\label{fig:fig3}
\end{figure}

$R_{0}$H is the contribution from OHE. $R_s$M is the contribution from AHE. The subscript xy indicates the resistivity of the sample has been measured by applying the current in the x direction and measuring the voltage in the y direction. $R_0$ is the ordinary Hall coefficient, $R_s$ is the AHE coefficient, M is the out-of-plane magnetization of the sample and $\rho_{THE}$ is the topological Hall resistivity of the sample. Therefore, the THE can be obtained by subtracting the other two contributions i.e., OHE and AHE  contributions from the total Hall resistivity of the sample. In figure \ref{fig:fig3}(a), the curve with blue circles depicts the Hall resistivity (AHE and THE) for our sample Ir-2 after subtracting the contribution from OHE by correcting the slope of the linear part of the total Hall resistivity 
($\rho_{xy}$). The OHE contribution is very small in our case which is shown in Fig. S5 (b) of the supplementary information. Further, the contribution from AHE is scaled with the $\rho_{xy}^{AHE + THE}$ by evaluating the AHE coefficient ($R_{s}$) which is plotted as the red line curve in figure \ref{fig:fig3} (a). The calculation can be summarized in the following\cite{raju2021colossal,swekis2019topological,tai2022distinguishing}, 
Equation \ref{eq:Rho} describes the total resistivity as a sum of three components. After subtracting the contribution from OHE, equation \ref{eq:Rho} becomes,
\begin{equation}
    \rho_{xy} = R_s M + \rho_{THE}
    \label{eq:Rho-Mod}
\end{equation}
At an applied magnetic field beyond the magnetic saturation, all the magnetic moment points in the direction of applied field and no topologically non-trivial spin textures are present in the system. Hence, $\rho_{THE}$ becomes zero and equation \ref{eq:Rho-Mod} becomes,
\begin{equation}
    \rho_{xy}^{sat} = R_s M_s
    \label{eq:Rho-Mod-2}
\end{equation}
Where $M_s$ is the saturation magnetization. Next, by putting the value of $R_s$ from equation \ref{eq:Rho-Mod-2} i.e., $\rho_{xy}^{sat}$/$M_s$ in equation \ref{eq:Rho-Mod} we extract the $\rho_{THE}$ as equation \ref{eq:Rho-Mod-3}.
\begin{equation}
    \rho_{THE} =\rho_{xy} - R_s M
    \label{eq:Rho-Mod-3}
\end{equation}
After subtracting the red line curve from the blue circle curve shown in figure \ref{fig:fig3} (a), we have found a hump kind of feature in both the field sweeps from +ve saturated state to -ve one as well as from -ve saturated state to +ve one, represented in the figure \ref{fig:fig3} (b) which is regarded as the contribution from THE. This shows the presence of chiral spin textures i.e. skyrmions in our system. The magnitude of THE in our sample is found to be 0.135 Nano Ohm-m. 
\section{Conclusion}
In summary, we have observed a canted state of magnetization reversal and presence of skyrmions in a Pt/Co PMA system with Ir spacer layer by varying only the Ir spacer layer thickness between the two Co layers. The MFM images show labyrinth kind of domains at the  demagnetized state and the presence of skyrmions is observed after applying external magnetic field to the samples. Further, we have observed a THE signal which indicates that the observed structures via MFM are of chiral spin textures. In this work we show that via the proper tuning of various energy along with inter-layer interactions (both dipolar and RKKY) it is possible to stabilize the skyrmions in the ferromagnet (e.g. Co) even when the thickness of the ferromagnet is well below the SRT regime. This paves the path to engineer such multilayer structure by varying the spacer layer thickness to have the presence of chiral spin textures such as skyrmions in the system. Further this structure may be investigated in future for current induced skyrmion motion by making devices out of it.
\section{Acknowledgement}
We acknowledge the department of atomic energy (DAE), Govt. of India, for proving funding and facilities to carry out the research work. We also thank the CEFIPRA (IFC/A/5808-1/2017/1395) project and  the Department of Science and Technology, Science and Engineering Research Board (Grant No. CRG/2021/001245) for their funding.
\section{References}
\bibliographystyle{iopart-num}
\bibliography{Reference}

\providecommand{\noopsort}[1]{}\providecommand{\singleletter}[1]{#1}%
\providecommand{\newblock}{}
\begin{thebibliography}{10}
\expandafter\ifx\csname url\endcsname\relax
  \def\url#1{{\tt #1}}\fi
\expandafter\ifx\csname urlprefix\endcsname\relax\def\urlprefix{URL }\fi
\providecommand{\eprint}[2][]{\url{#2}}

\bibitem{luo2018reconfigurable}
Luo S, Song M, Li X, Zhang Y, Hong J, Yang X, Zou X, Xu N and You L 2018 {\em
  Nano letters\/} {\bf 18} 1180--1184

\bibitem{dash2023device}
Dash A, Ojha B, Mohanty S, Moharana A~K and Bedanta S 2023 {\em
  Nanotechnology\/} {\bf 34} 185001

\bibitem{zhang2015magnetic}
Zhang X, Ezawa M and Zhou Y 2015 {\em Scientific reports\/} {\bf 5} 9400

\bibitem{skyrme1961non}
Skyrme T~H~R 1961 {\em Proceedings of the Royal Society of London. Series A.
  Mathematical and Physical Sciences\/} {\bf 260} 127--138

\bibitem{bogdanov1989thermodynamically}
Bogdanov A~N and Yablonskii D 1989 {\em Zh. Eksp. Teor. Fiz\/} {\bf 95} 178

\bibitem{bogdanov1994thermodynamically}
Bogdanov A and Hubert A 1994 {\em Journal of magnetism and magnetic
  materials\/} {\bf 138} 255--269

\bibitem{bogdanov2020physical}
Bogdanov A~N and Panagopoulos C 2020 {\em Nature Reviews Physics\/} {\bf 2}
  492--498

\bibitem{muhlbauer2009skyrmion}
Muhlbauer S, Binz B, Jonietz F, Pfleiderer C, Rosch A, Neubauer A, Georgii R
  and Boni P 2009 {\em Science\/} {\bf 323} 915--919

\bibitem{leonov2016properties}
Leonov A, Monchesky T, Romming N, Kubetzka A, Bogdanov A and Wiesendanger R
  2016 {\em New Journal of Physics\/} {\bf 18} 065003

\bibitem{moreau2016additive}
Moreau-Luchaire C, Moutafis C, Reyren N, Sampaio J, Vaz C, Van~Horne N,
  Bouzehouane K, Garcia K, Deranlot C, Warnicke P {\em et~al.\/} 2016 {\em
  Nature nanotechnology\/} {\bf 11} 444--448

\bibitem{boulle2016room}
Boulle O, Vogel J, Yang H, Pizzini S, de~Souza~Chaves D, Locatelli A,
  Mente{\c{s}} T~O, Sala A, Buda-Prejbeanu L~D, Klein O {\em et~al.\/} 2016
  {\em Nature nanotechnology\/} {\bf 11} 449--454

\bibitem{rohart2013skyrmion}
Rohart S and Thiaville A 2013 {\em Physical Review B\/} {\bf 88} 184422

\bibitem{von2017enhanced}
von Malottki S, Dup{\'e} B, Bessarab P~F, Delin A and Heinze S 2017 {\em
  Scientific reports\/} {\bf 7} 12299

\bibitem{ezawa2010giant}
Ezawa M 2010 {\em Physical review letters\/} {\bf 105} 197202

\bibitem{fert2013skyrmions}
Fert A, Cros V and Sampaio J 2013 {\em Nature nanotechnology\/} {\bf 8}
  152--156

\bibitem{fert2017magnetic}
Fert A, Reyren N and Cros V 2017 {\em Nature Reviews Materials\/} {\bf 2} 1--15

\bibitem{nagaosa2013topological}
Nagaosa N and Tokura Y 2013 {\em Nature nanotechnology\/} {\bf 8} 899--911

\bibitem{tomasello2014strategy}
Tomasello R, Martinez E, Zivieri R, Torres L, Carpentieri M and Finocchio G
  2014 {\em Scientific reports\/} {\bf 4} 1--7

\bibitem{zhang2015skyrmion}
Zhang X, Zhao G, Fangohr H, Liu J~P, Xia W, Xia J and Morvan F 2015 {\em
  Scientific reports\/} {\bf 5} 7643

\bibitem{sampaio2013nucleation}
Sampaio J, Cros V, Rohart S, Thiaville A and Fert A 2013 {\em Nature
  nanotechnology\/} {\bf 8} 839--844

\bibitem{ojha2023driving}
Ojha B, Mallick S, Panigrahy S, Sharma M, Thiaville A, Rohart S and Bedanta S
  2023 {\em Physica Scripta\/} {\bf 98} 035819

\bibitem{chen2022controllable}
Chen R, Cui Q, Han L, Xue X, Liang J, Bai H, Zhou Y, Pan F, Yang H and Song C
  2022 {\em Advanced Functional Materials\/} {\bf 32} 2111906

\bibitem{dohi2019formation}
Dohi T, DuttaGupta S, Fukami S and Ohno H 2019 {\em Nature communications\/}
  {\bf 10} 5153

\bibitem{chen2020realization}
Chen R, Gao Y, Zhang X, Zhang R, Yin S, Chen X, Zhou X, Zhou Y, Xia J, Zhou Y
  {\em et~al.\/} 2020 {\em Nano letters\/} {\bf 20} 3299--3305

\bibitem{legrand2020room}
Legrand W, Maccariello D, Ajejas F, Collin S, Vecchiola A, Bouzehouane K,
  Reyren N, Cros V and Fert A 2020 {\em Nature materials\/} {\bf 19} 34--42

\bibitem{raju2019evolution}
Raju M, Yagil A, Soumyanarayanan A, Tan A~K, Almoalem A, Ma F, Auslaender O and
  Panagopoulos C 2019 {\em Nature communications\/} {\bf 10} 696

\bibitem{bruno1991oscillatory}
Bruno e~P and Chappert C 1991 {\em Physical review letters\/} {\bf 67} 1602

\bibitem{bruno1992ruderman}
Bruno P and Chappert C 1992 {\em Physical Review B\/} {\bf 46} 261

\bibitem{mohanty2022magnetization}
Mohanty S, Sharma M, Moharana A~K, Ojha B, Pandey E, Singh B~B and Bedanta S
  2022 {\em Jom\/} {\bf 74} 2319--2327

\bibitem{duine2018synthetic}
Duine R, Lee K~J, Parkin S~S and Stiles M~D 2018 {\em Nature physics\/} {\bf
  14} 217--219

\bibitem{salikhov2022control}
Salikhov R, Samad F, Arekapudi S~S~P~K, Ehrler R, Lindner J, Kiselev N~S and
  Hellwig O 2022 {\em Physical Review B\/} {\bf 106} 054404

\bibitem{palotas2017spin}
Palot{\'a}s K, R{\'o}zsa L, Simon E, Udvardi L and Szunyogh L 2017 {\em
  Physical Review B\/} {\bf 96} 024410

\bibitem{juge2022skyrmions}
Juge R, Sisodia N, Larra{\~n}aga J~U, Zhang Q, Pham V~T, Rana K~G, Sarpi B,
  Mille N, Stanescu S, Belkhou R {\em et~al.\/} 2022 {\em Nature
  Communications\/} {\bf 13} 4807

\bibitem{he2018evolution}
He M, Li G, Zhu Z, Zhang Y, Peng L, Li R, Li J, Wei H, Zhao T, Zhang X~G {\em
  et~al.\/} 2018 {\em Physical Review B\/} {\bf 97} 174419

\bibitem{raju2021colossal}
Raju M, Petrovi{\'c} A, Yagil A, Denisov K, Duong N, G{\"o}bel B,
  {\c{S}}a{\c{s}}{\i}o{\u{g}}lu E, Auslaender O, Mertig I, Rozhansky I {\em
  et~al.\/} 2021 {\em Nature Communications\/} {\bf 12} 2758

\bibitem{soumyanarayanan2017tunable}
Soumyanarayanan A, Raju M, Gonzalez~Oyarce A, Tan A~K, Im M~Y, Petrovi{\'c}
  A~P, Ho P, Khoo K, Tran M, Gan C {\em et~al.\/} 2017 {\em Nature materials\/}
  {\bf 16} 898--904

\bibitem{chen2020magnetic}
Chen R, Zhang R, Zhou Y, Bai H, Pan F and Song C 2020 {\em Applied Physics
  Letters\/} {\bf 116}

\bibitem{sivakumar2020topological}
Sivakumar P~K, Gobel B, Lesne E, Markou A, Gidugu J, Taylor J~M, Deniz H, Jena
  J, Felser C, Mertig I {\em et~al.\/} 2020 {\em ACS nano\/} {\bf 14}
  13463--13469

\bibitem{kimbell2022challenges}
Kimbell G, Kim C, Wu W, Cuoco M and Robinson J~W 2022 {\em Communications
  Materials\/} {\bf 3} 19

\bibitem{swekis2019topological}
Swekis P, Markou A, Kriegner D, Gayles J, Schlitz R, Schnelle W, Goennenwein
  S~T and Felser C 2019 {\em Physical Review Materials\/} {\bf 3} 013001

\bibitem{tai2022distinguishing}
Tai L, Dai B, Li J, Huang H, Chong S~K, Wong K~L, Zhang H, Zhang P, Deng P,
  Eckberg C {\em et~al.\/} 2022 {\em ACS nano\/} {\bf 16} 17336--17346

\end{thebibliography}
\end{document}